\documentclass{article}
\usepackage{spconf,amsmath,graphicx}
\newcommand{\etal}{\textit{et al.}}
\usepackage{cite}
\usepackage[linesnumbered,ruled,vlined]{algorithm2e}
\usepackage{multirow}
\usepackage{caption,subcaption}
\usepackage{xcolor}

\title{Deep No-Reference Tone Mapped Image Quality Assessment}

\name{Chandra Sekhar Ravuri$^1$, Rajesh Sureddi$^2$, Sathya Veera Reddy Dendi$^2$,}

\secondlinename{Shanmuganathan Raman$^1$, Sumohana S. Channappayya$^2$}

\address{$^1$Department of Electrical Engineering, Indian Institute of Technology Gandhinagar, India. \\ 
$^2$Department of Electrical Engineering, Indian Institute of Technology Hyderabad, India. 
}

\begin{document}
%
\maketitle
\begin{abstract}
The process of rendering high dynamic range (HDR) images to be viewed on conventional displays is called tone mapping. However, tone mapping introduces distortions in the final image which may lead to visual displeasure. To quantify these distortions, we introduce a novel no-reference quality assessment technique for these tone mapped images. This technique is composed of two stages. In the first stage, we  employ a convolutional neural network (CNN) to generate quality aware maps (also known as distortion maps) from tone mapped images by training it with the ground truth distortion maps. In the second stage, we model the normalized image and distortion maps using an Asymmetric Generalized Gaussian Distribution (AGGD). The parameters of the AGGD model are then used to estimate the quality score using support vector regression (SVR). We show that the proposed technique delivers competitive performance relative to the state-of-the-art techniques. The novelty of this work is its ability to visualize various distortions as quality maps (distortion maps), especially in the no-reference setting, and to use these maps as features to estimate the quality score of tone mapped images.
\end{abstract}
\section{Introduction}
\label{sec:intro}
The real world scene contains a very high dynamic range (HDR) in the radiance space and it is not possible to capture all the intensity levels in the scene. Generally, we capture in the order of 256 levels per channel using normal cameras, but with the advancement in camera technology, we are able to capture higher intensity range in the order of 10,000 levels. These images are called HDR images. HDR images give visual pleasure and details while capturing the scene, which almost looks like the real world. To visualize HDR images, we require special displays which are very expensive. In order to display the HDR images on normal displays, we need to map from high dynamic range to a lower dynamic range (LDR) in the order of 256 levels. This process is called tone mapping. Even though the image looks pristine in HDR, we may lose some visual information because of this compression in the intensity dynamic range. The tone mapping operator should be decided on how well it can preserve all the perceptual information present in the HDR image while obtaining the tone mapped image. With this requirement, we need to assess the perceptual quality of a tone mapped image. The best way to assess the quality of these images is via human  opinion scores. However, collecting human scores is an expensive and time-consuming process.  This drawback of subjective quality assessment necessitates the design of objective quality assessment algorithms. In this context, algorithms like the tone mapped image quality index (TMQI) in \cite{yeganeh2013objective} were implemented with the requirement of an original HDR image as reference to compare with the tone mapped image. However, it is not always possible to store HDR images which are memory intensive. Hence, there is a requirement for quality assessment metrics that do not require a reference - also called as no-reference image quality assessment (NRIQA). One of the major distortions that might occur in the tone mapping process is contrast distortion given the HDR images contain only static objects. We have addressed this problem in the proposed work. In this work, we propose a CNN based NRIQA algorithm for measuring tone mapped image quality. In brief, the proposed approach has two stages. In the first stage, a test image is mapped to the quality-aware domain (distortion maps) and in the next stage, the quality-aware representations are mapped to a single perceptual quality score. 

We briefly review the existing NRIQA techniques of both LDR and HDR images.
NIQE \cite{mittal2013making} is an opinion unaware (i.e., does not use subjective scores) and distortion unaware (i.e., does not rely on distortion type) NRIQA technique. ILNIQE~\cite{zhang2015feature} is a variant of NIQE based on multiple cues.
The DIIVINE \cite{moorthy2011blind} framework identifies the type of the distortion first and assesses the image quality next. 
Another blind image quality assessment technique by Gu \etal~in~\cite{gu2014deep} uses deep neural networks (DNN) with NSS features as input and image quality assessment was treated as a classification task. 
Kang \etal~in~\cite{kang2014convolutional} proposed a NRIQA technique using CNN.
HDR images have very few NRIQA techniques, unlike LDR images. Gu \etal~in~\cite{gu2016blind} proposed an NRIQA technique for tone mapped images using naturalness and structural information of the tone mapped image to predict the quality of the tone mapped image. HIGRADE \cite{HIGRADE} is NRIQA algorithm proposed based on gradient scene-statistics, defined in the LAB color space. DRIIQA is a full reference image quality assessment (FRIQA) technique by Aydin \etal~in~\cite{aydin2008dynamic} which generates a quality map of the test image with respective to its reference image. These maps are called distortion maps, which localize the distortion in space. However, this technique does not give a single quality score of a test image. After studying the existing techniques, we propose a new framework for NRIQA of tone mapped images by generating distortion maps in~\cite{aydin2008dynamic} using CNN and extracting features from them to map to a single quality score. Our algorithm will be explained in the following sections.
\section{Proposed Approach}
\label{sec:my_work}
We follow the work in~\cite{aydin2008dynamic} where they generated distortion maps using both HDR and the tone mapped image to visualize different distortions, namely amplification of invisible contrast, loss of visible contrast, and reversal of visible contrast in both high-pass and low-pass bands. The problem that we address in this work is to remove the dependency on the original HDR image~\cite{aydin2008dynamic} i.e., transition from full-reference quality assessment to no-reference quality assessment. 
To extend this work, the distortion maps are used as features to calculate a single quality score for a tone mapped image.
In brief, the proposed metric is composed of two stages. In the first stage, we generate distortion maps using the proposed convolutional neural network. In the second stage, we model the distortion map intensities using an AGGD and use the model parameters to predict the overall quality score. 
\subsection{Dataset Generation and Description}
\label{sec:dataset}
To train the proposed RcNet shown in Fig.~\ref{fig:RcNet}, we follow the work in~\cite{aydin2008dynamic} to generate ground truth distortion maps by using HDR images and corresponding tone mapped images (a FRIQA approach). Our efforts to find datasets which contain both the original HDR images and corresponding tone mapped images resulted only one database~\cite{yeganeh2013objective}. Alternatively, we identified and collected HDR images from different sources including~\cite{xiao2002high, kronander2014unified,fairchild2007hdr,azimi2014evaluating}. We do not have any information about the tone-mapped images for the collected HDR images. Hence, we selected five popular tone-mapping algorithms~\cite{reinhard2002photographic,fattal2002gradient,ward1997visibility,durand2002fast,raman2009bilateral} and operated on the collected HDR images. We generated 5,325 tone mapped images with their default parameters of tone mapping operators using the HDR toolbox~\cite{Banterle:2017}. From the collected 5,325 images, 3,460 images are used for training, 1,000 images are used for testing, and remaining 865 images are used for validation.

While these tone mapped images act as the input to our CNN, we do not have corresponding labels yet. To train the CNN, which will be explained in detail in the next section, we need ground truth distortion maps (labels). To achieve this, we generated distortion maps using the algorithm proposed by Aydin \etal~in~\cite{aydin2008dynamic}. The motivation behind using this algorithm is that when tone mapping an HDR image, the most likely distortion resulting from the process is contrast distortion. These distortions are clearly captured by this full reference algorithm. We provided the collected HDR images and the corresponding tone mapped images to this algorithm and generated six distortion maps per tone mapped image. We used these distortion maps as labels to train the CNN.
\subsection{Training Procedure}
\label{training}
\begin{figure}[h]
\centering
\includegraphics[width=\linewidth]{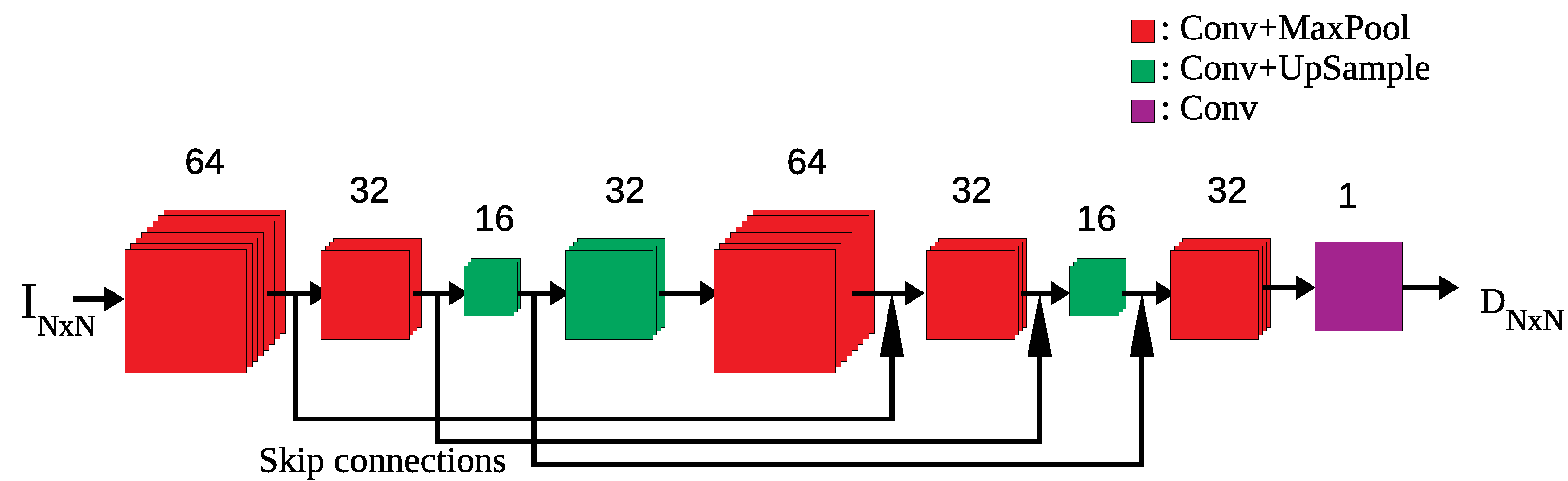}
\caption{Network architecture of the proposed RcNet}
\label{fig:RcNet}
\end{figure}

We trained the neural network architecture shown in Fig.~\ref{fig:RcNet}. The numbers in the architecture indicate the depth of convolution maps. It accepts the tone mapped image (I) as input and distortion map (D) as label. As mentioned in the previous section, for each tone mapped image, there are six corresponding distortion maps. Hence, we trained a different model for each distortion map with the same architecture. There are two options to train the network: one is to compromise on computations and train the network for the large input size, where the size of the image may not be same and also may have different aspect ratios. Another is to split the data into patches and train the network. We adopted the second option to train our network. A natural question that arises due to this alternative is about the optimal patch size. To answer this question, we trained our network with overlapping patches of three different sizes: $64 \times 64$, $128 \times 128$, and $256 \times 256$. A patch size of $128 \times 128$ gave us the best performance and is chosen in this work. While designing this network, we imposed the following constraints: a) low complexity, b) avoid over-fitting, c) use initial layer features in final layers with skip connections. We have used pooling, up-sampling, and dropouts at appropriate layers in proposed network to make it more robust.

\begin{figure}[!htbp]
\centering
\includegraphics[width=5cm,height=3cm]{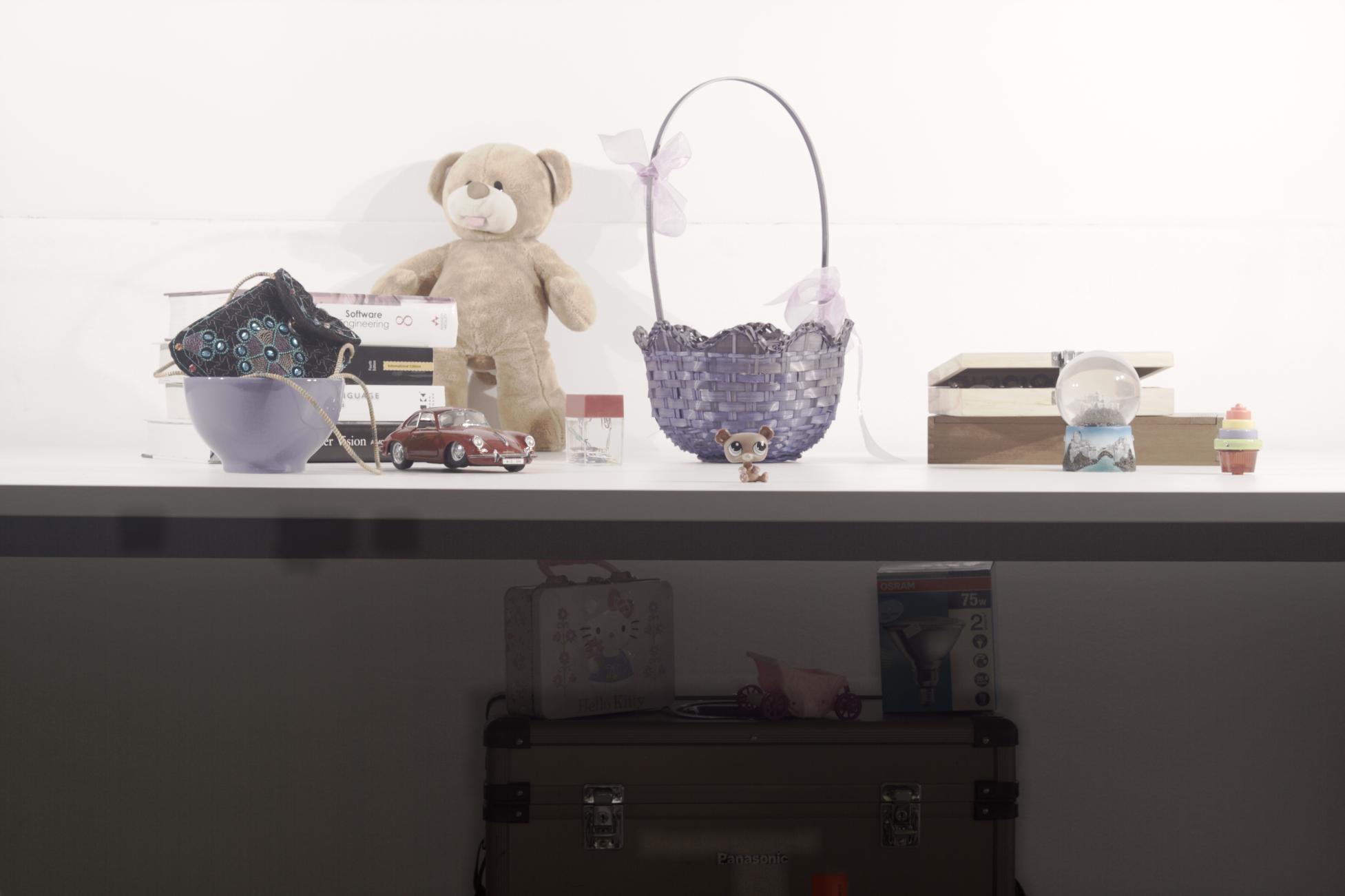}
\caption{Tone-mapped image from test dataset~\cite{karaduzovic2017multi}.}
\label{fig:test_dist_tone}
\end{figure}

\begin{figure*}[!htbp]
\centering
\begin{minipage}[b]{0.17\linewidth}
\centering
\includegraphics[width=\textwidth]{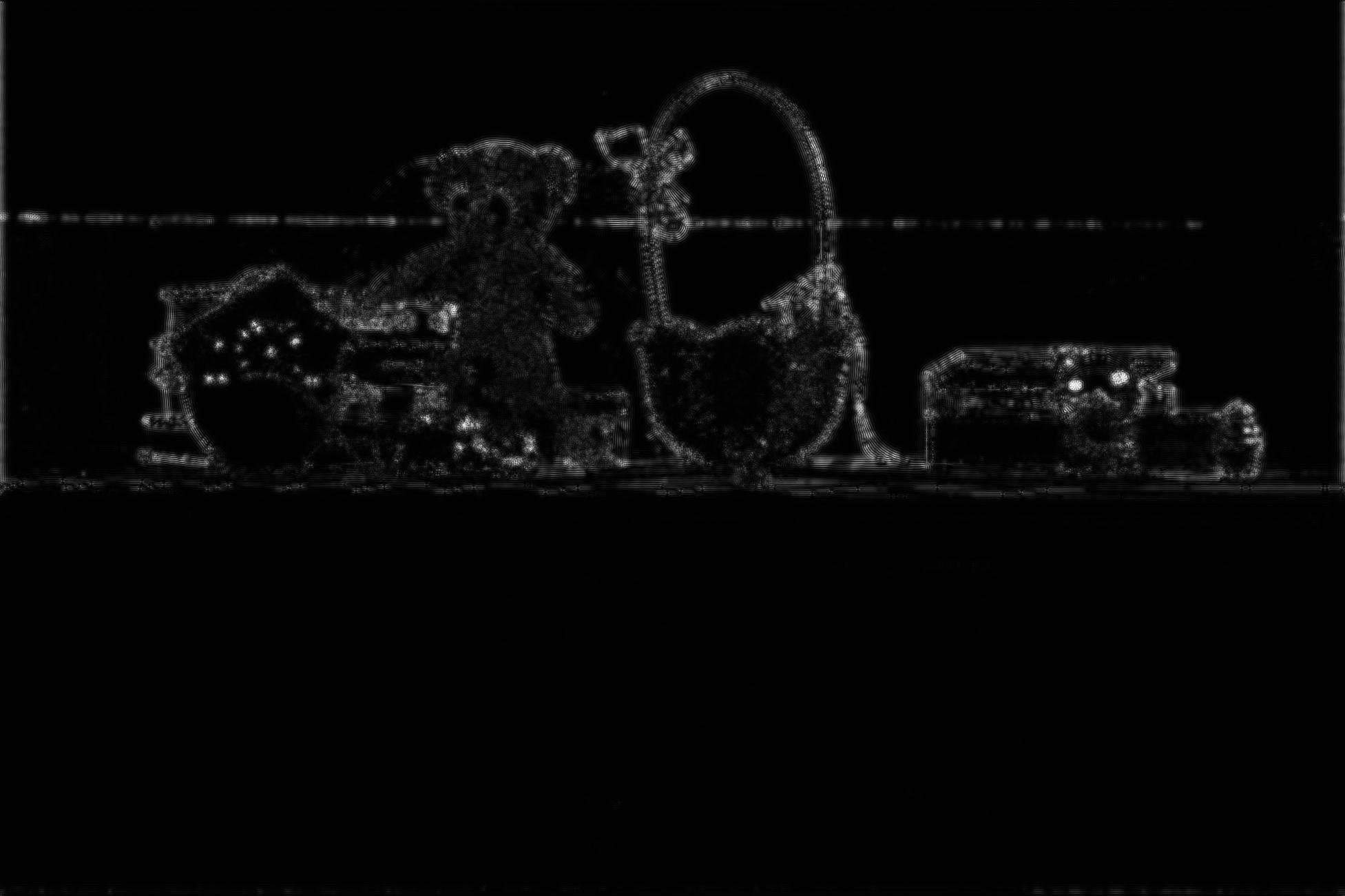}
{\center  (a) }
\end{minipage}%
%
\begin{minipage}[b]{0.17\linewidth}
\centering
\includegraphics[width=\textwidth]{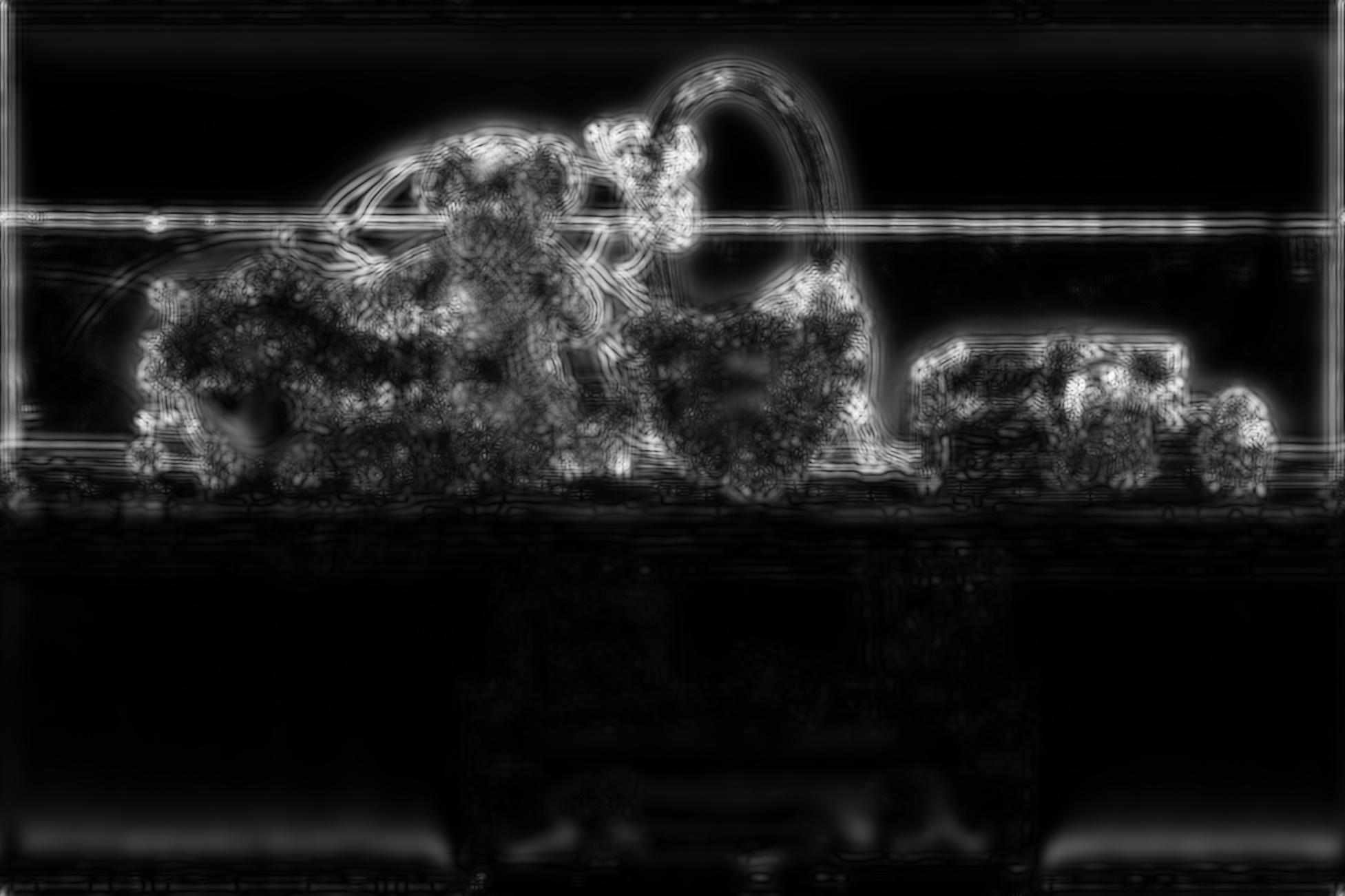}
{\center  (b) }
\end{minipage}%
%
%
\begin{minipage}[b]{0.17\linewidth}
\centering
\includegraphics[width=\textwidth]{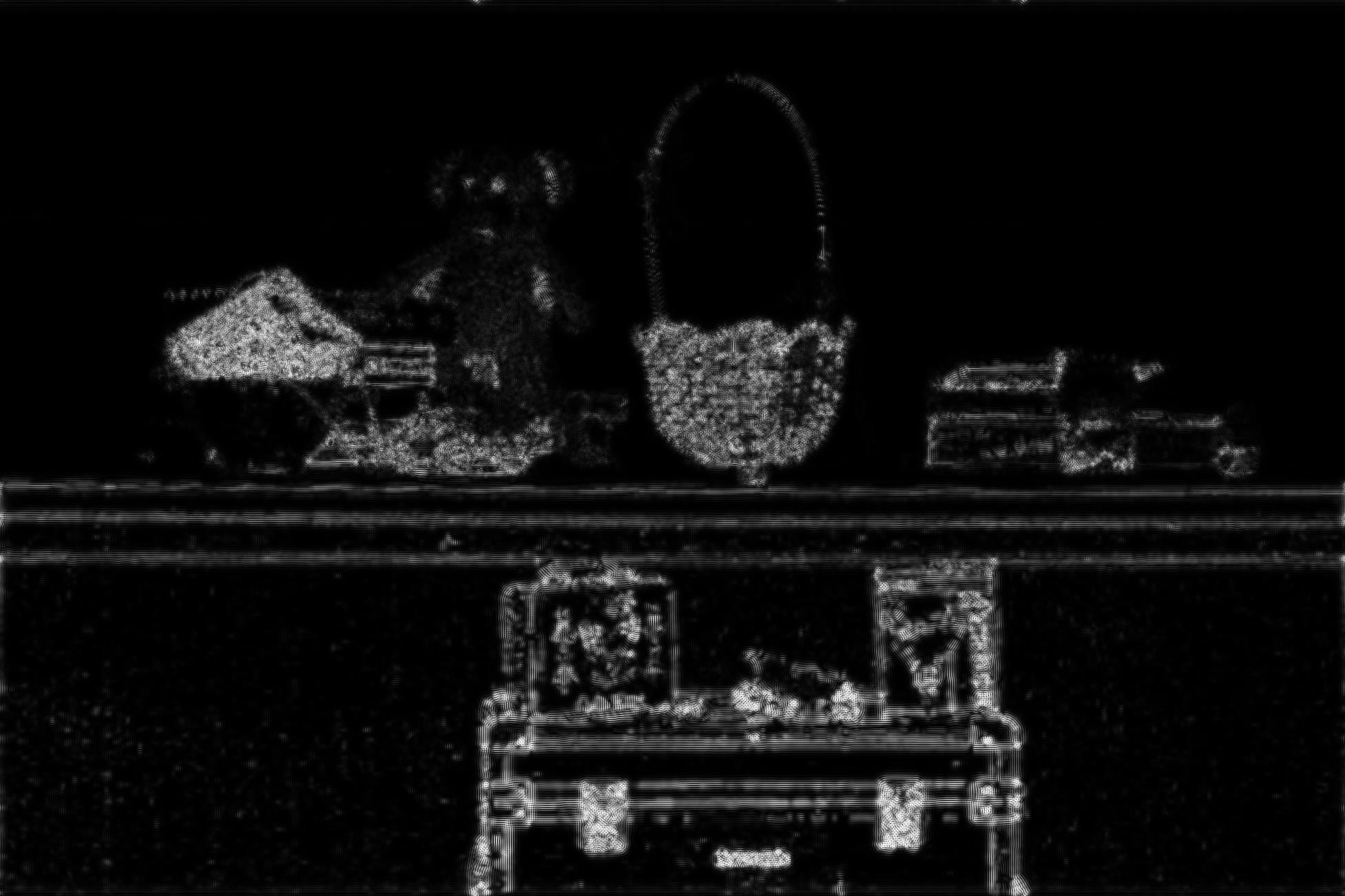}
{\center  (c) }
\end{minipage}%
%
%
\begin{minipage}[b]{0.17\linewidth}
\centering
\includegraphics[width=\textwidth]{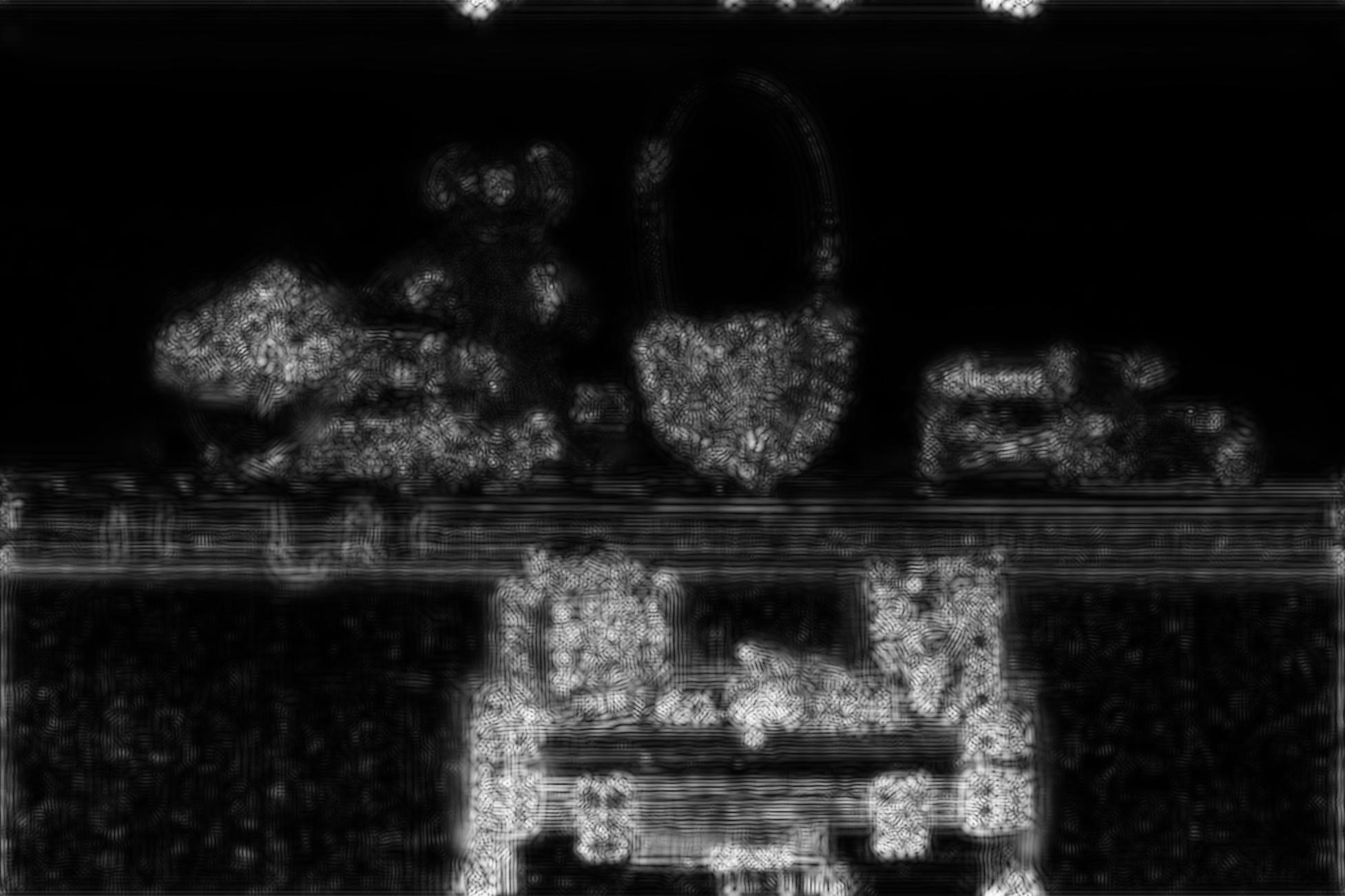}
{\center  (d) }
\end{minipage}%
\begin{minipage}[b]{0.17\linewidth}
\centering
\includegraphics[width=\textwidth]{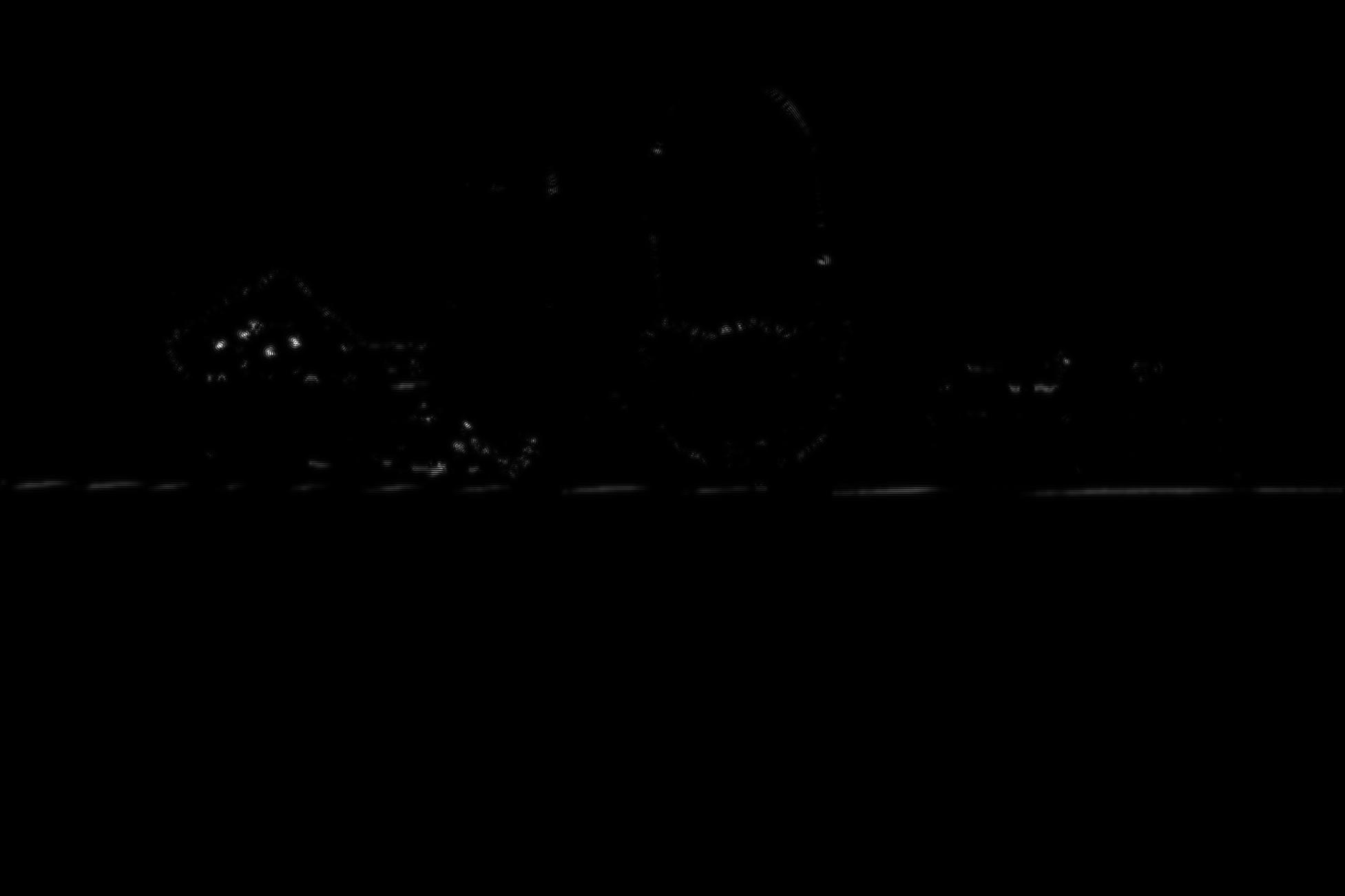}
{\center  (e) }
\end{minipage}%
\begin{minipage}[b]{0.17\linewidth}
\centering
\includegraphics[width=\textwidth]{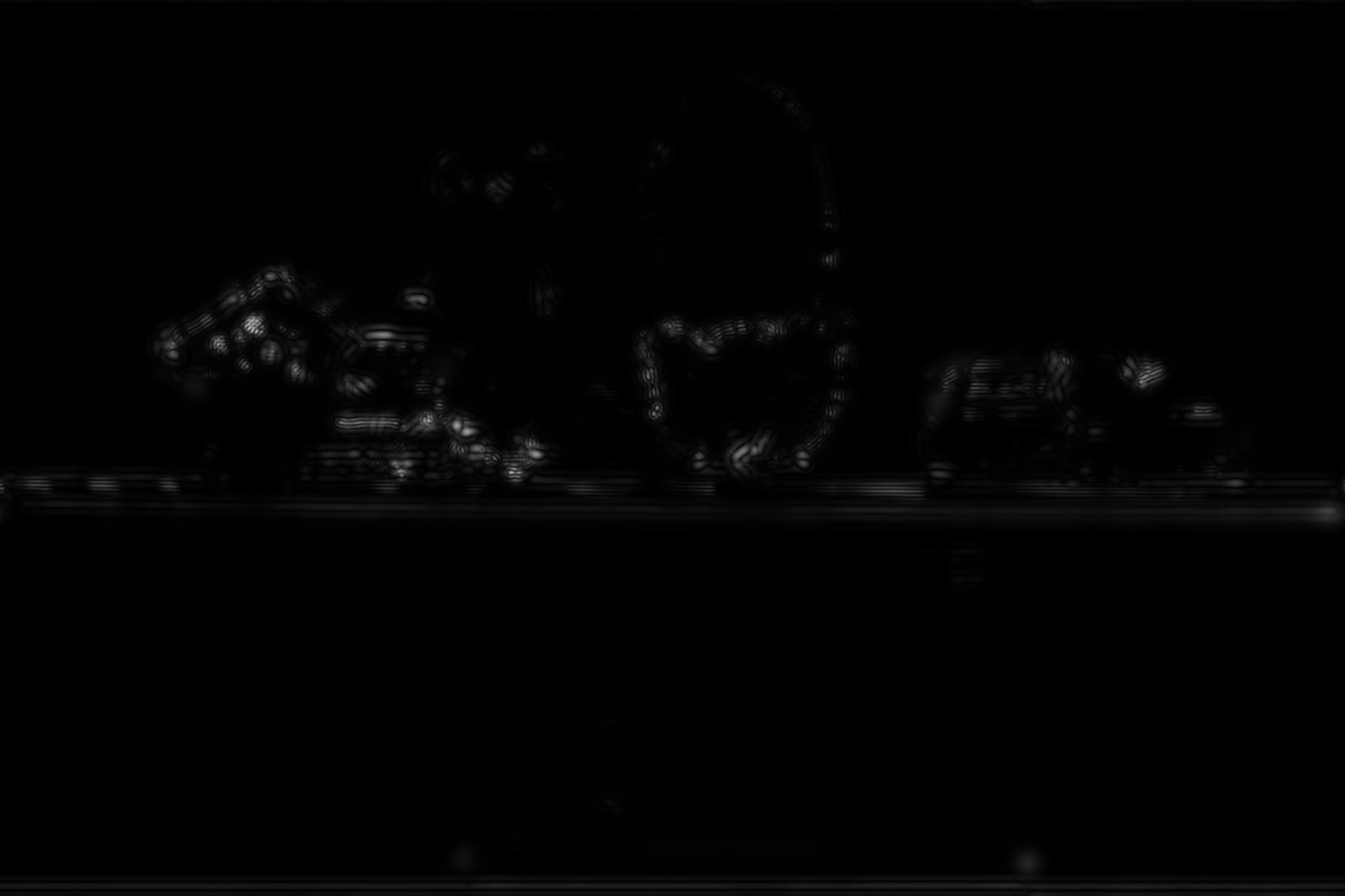}
{\center  (f) }
\end{minipage}%
\vfill
\begin{minipage}[b]{0.17\linewidth}
\centering
\includegraphics[width=\textwidth]{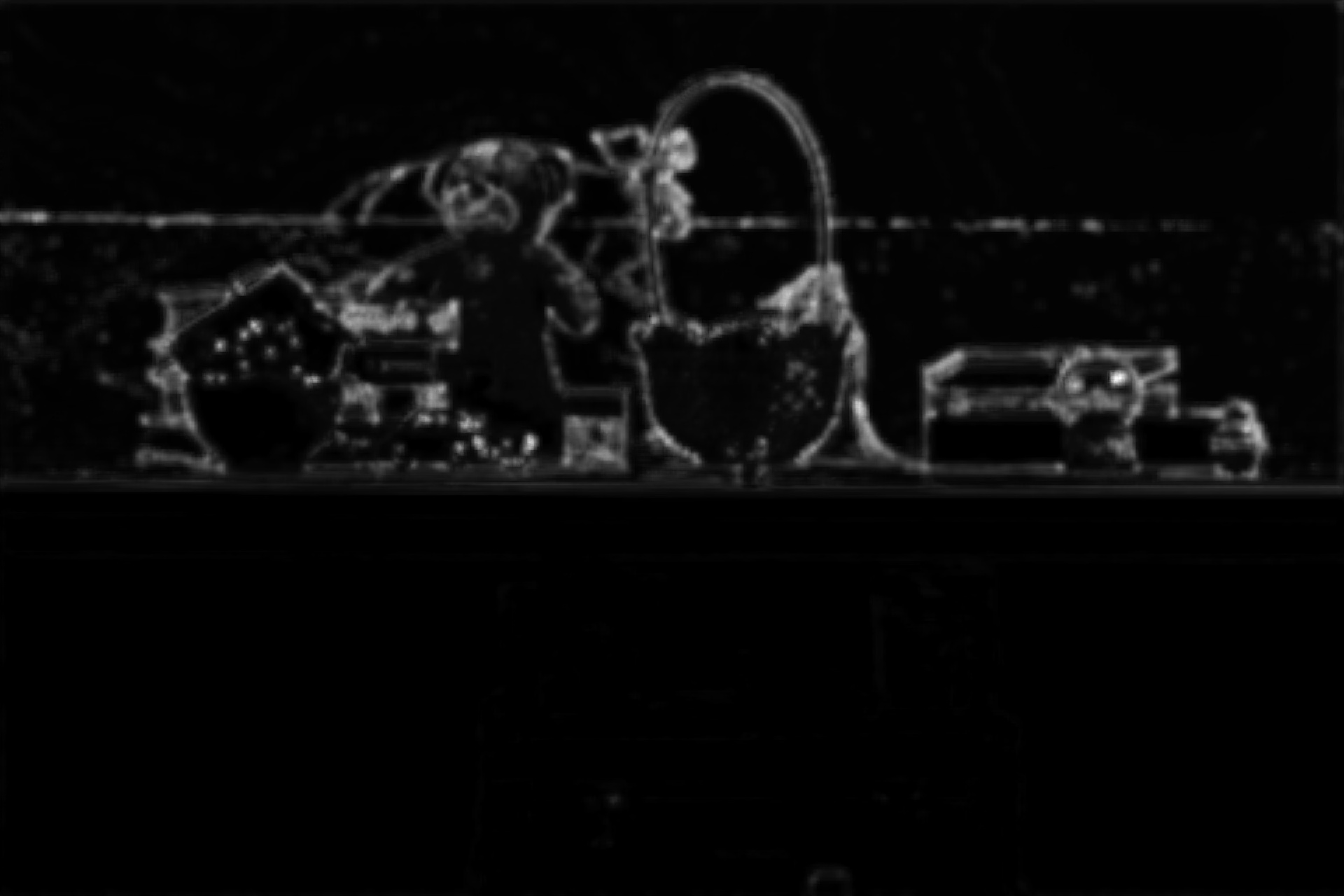}
{\center  (g) }
\end{minipage}%
\begin{minipage}[b]{0.17\linewidth}
\centering
\includegraphics[width=\textwidth]{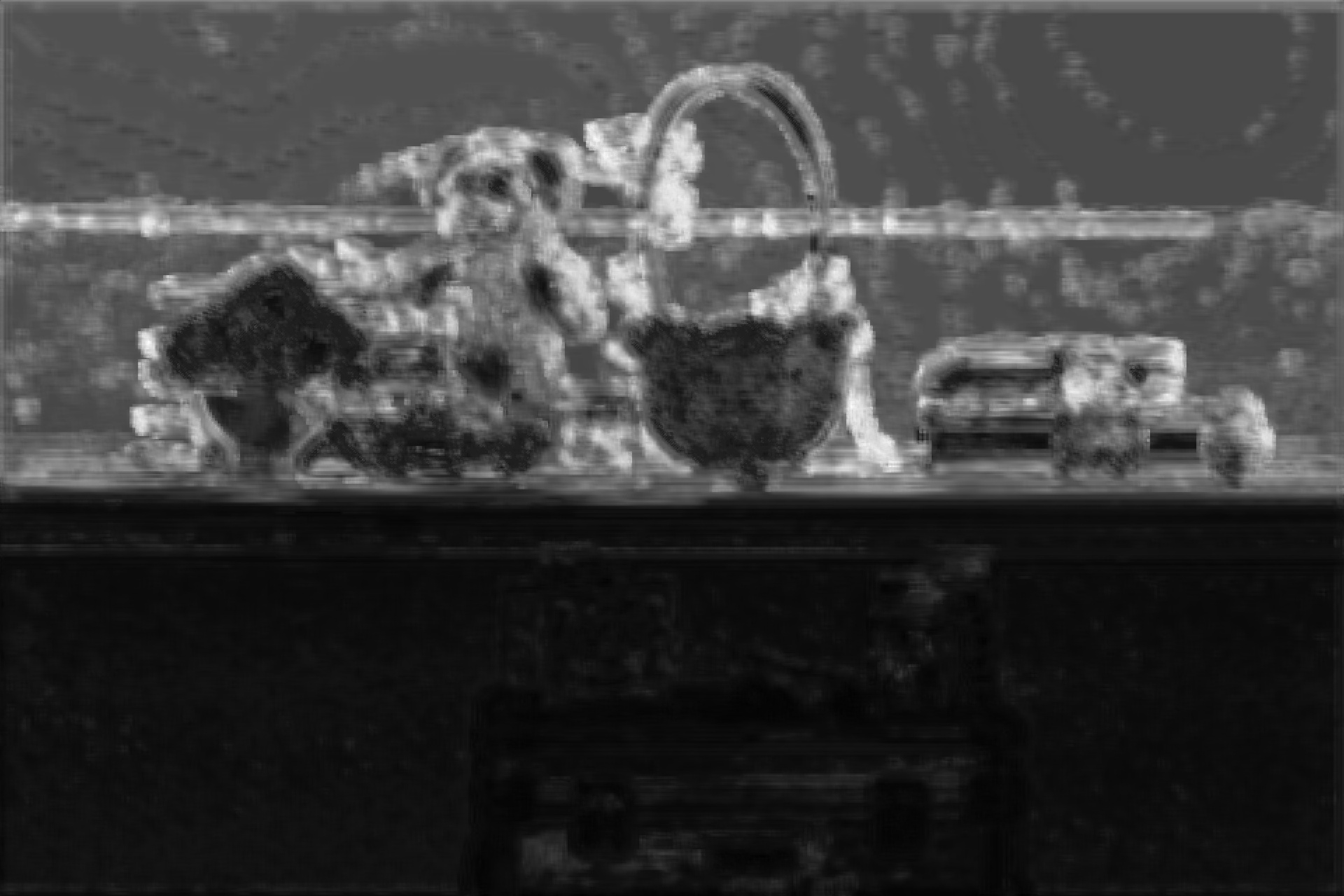}
{\center  (h) }
\end{minipage}%
\begin{minipage}[b]{0.17\linewidth}
\centering
\includegraphics[width=\textwidth]{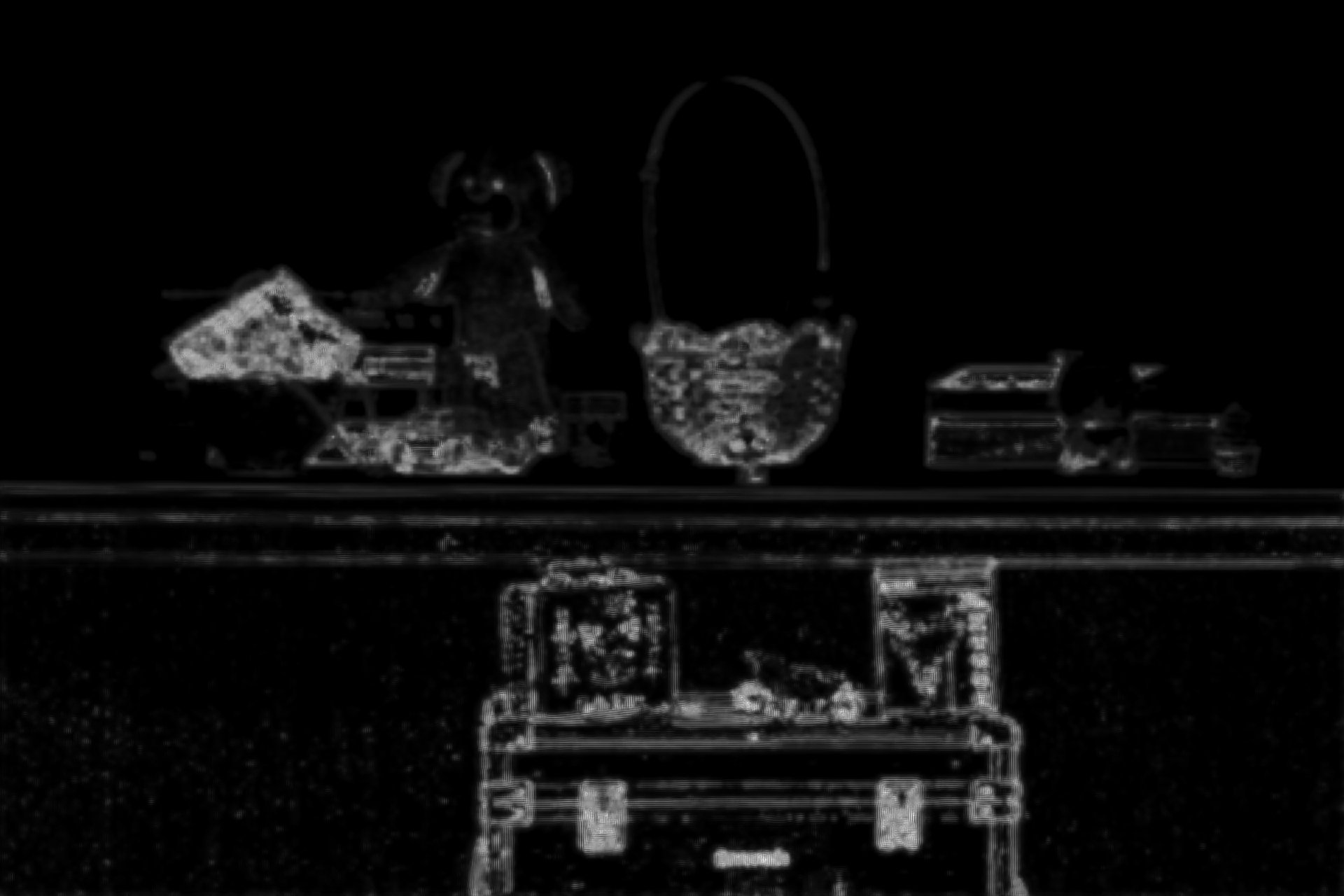}
{\center  (i) }
\end{minipage}%
\begin{minipage}[b]{0.17\linewidth}
\centering
\includegraphics[width=\textwidth]{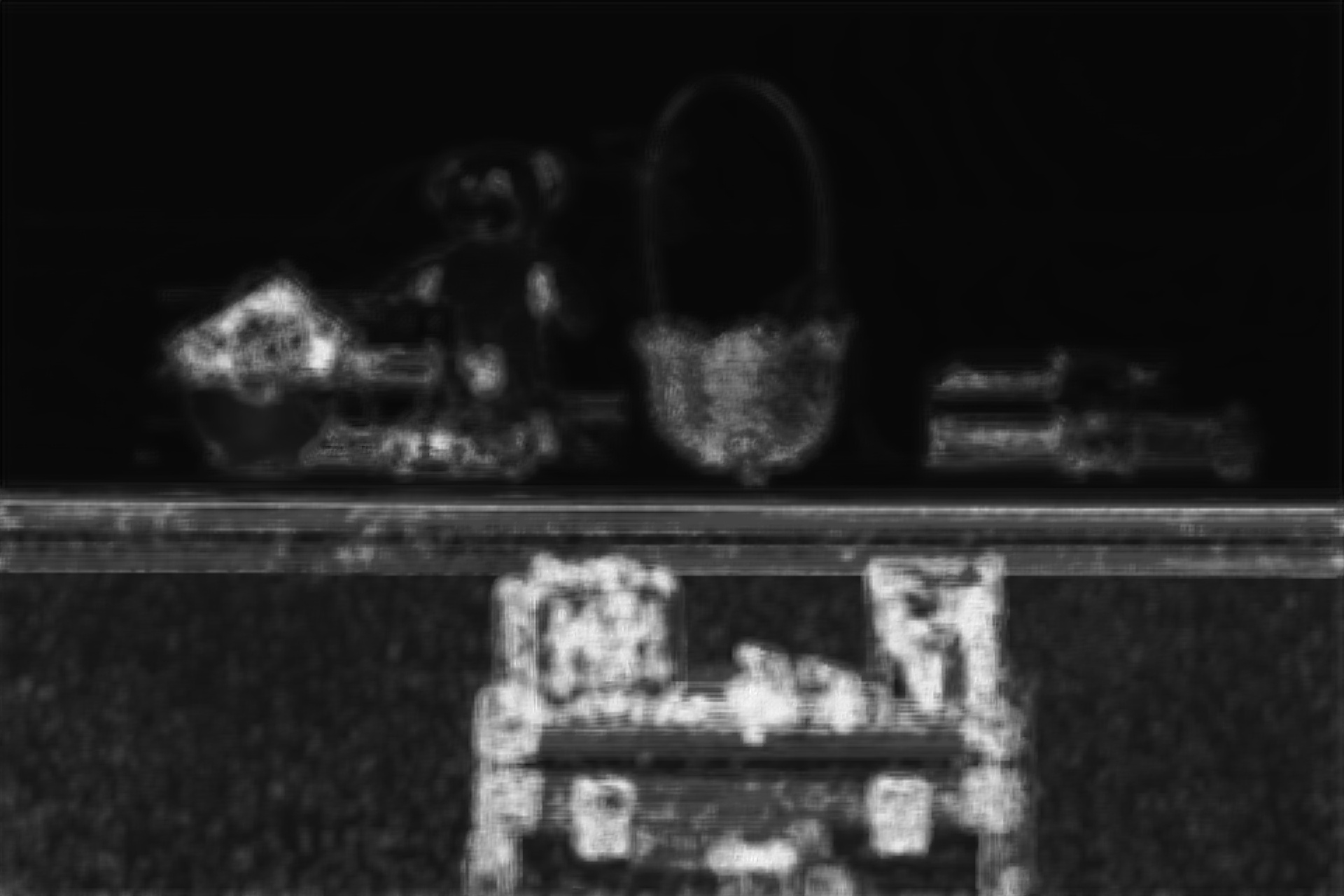}
{\center  (j) }
\end{minipage}%
%
%
\begin{minipage}[b]{0.17\linewidth}
\centering
\includegraphics[width=\textwidth]{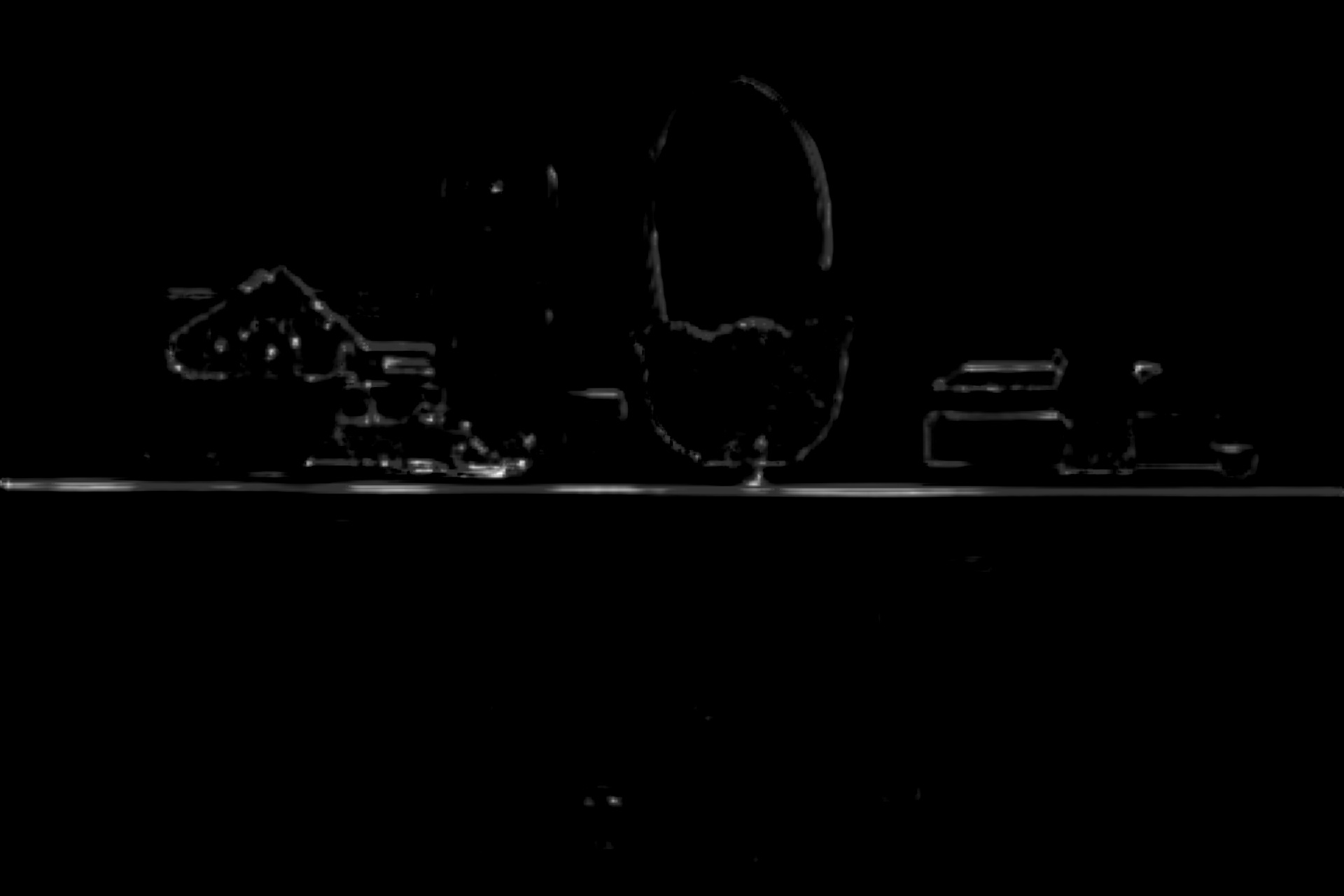}
{\center  (k) }
\end{minipage}%
\begin{minipage}[b]{0.17\linewidth}
\centering
\includegraphics[width=\textwidth]{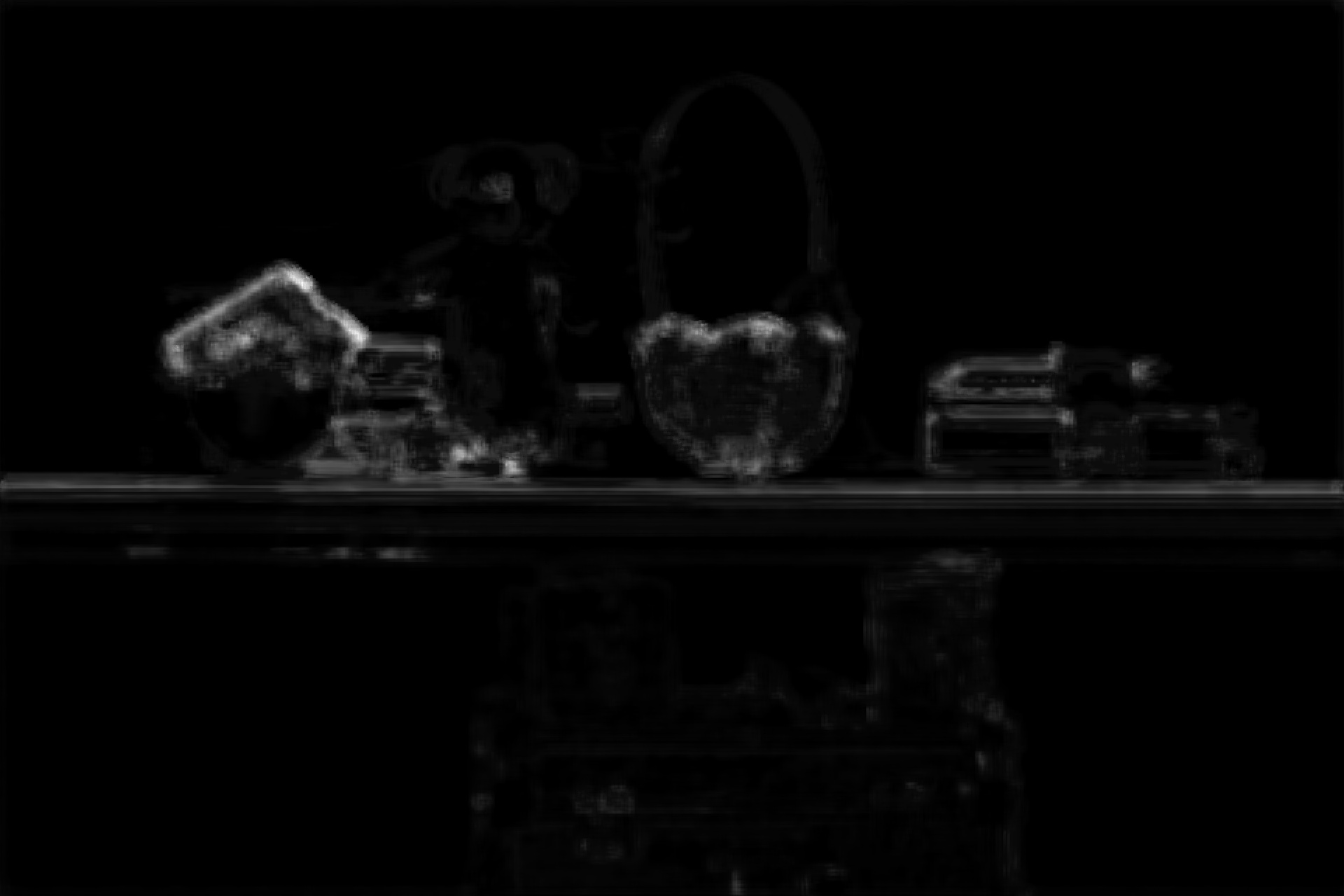}
{\center  (l) }
\end{minipage}%
\caption{(a) - (f) represent ground truth distortion maps and (g) - (l) represent predicted distortion maps with proposed RcNet.}
\label{fig:compare_dist_maps}
\end{figure*}
\subsection{Stage-I: Estimation of Distortion-maps}
In this stage, we explain the details about the generation of quality maps or distortion maps. The distortion maps are the representations of the amount of distortion present at each pixel introduced by the tone mapping operation. As described earlier, we generate distortion maps using a convolutional neural network which will predict the pixel wise distortions after sufficient training. 
After experimenting with several deep network architectures (including auto encoders) we found that the RcNet architecture in Fig.~\ref{fig:RcNet} to be best suited for our problem.
The first stage of the quality metric is to generate the distortion maps using the trained models. The results of applying the models on an image as shown in Fig. \ref{fig:test_dist_tone} from the dataset~\cite{karaduzovic2017multi} are shown in Fig. \ref{fig:compare_dist_maps}. 

Fig. \ref{fig:compare_dist_maps} (a) represents the ground truth distortion map of amplification of contrast in the high-pass band, and Fig. \ref{fig:compare_dist_maps} (g) represents the corresponding predicted distortion map using RcNet as shown in Fig. \ref{fig:RcNet}. Similarly, Fig. \ref{fig:compare_dist_maps}(b), Fig. \ref{fig:compare_dist_maps}(c), Fig. \ref{fig:compare_dist_maps}(d), Fig. \ref{fig:compare_dist_maps}(e), and Fig. \ref{fig:compare_dist_maps}(f) represent the ground truth distortion maps of amplification of contrast in the low-pass band, loss of contrast in the high-pass band, loss of contrast in the low-pass band, reversal of contrast in the high-pass band, and reversal of contrast in the low-pass band, respectively. Fig. \ref{fig:compare_dist_maps}(h), Fig. \ref{fig:compare_dist_maps}(i), Fig. \ref{fig:compare_dist_maps}(j), Fig. \ref{fig:compare_dist_maps}(k), and Fig. \ref{fig:compare_dist_maps}(l) represent the corresponding predicted distortion maps using RcNet in Fig.~\ref{fig:RcNet}. The similarity between the generated and ground truth distortion maps provides a qualitative illustration of our network's ability for distortion map generation.
\subsection{Stage-II: Estimation of Quality Scores}
\begin{figure}[!htbp]
\centering
\includegraphics[width=5.5cm,height=4cm]{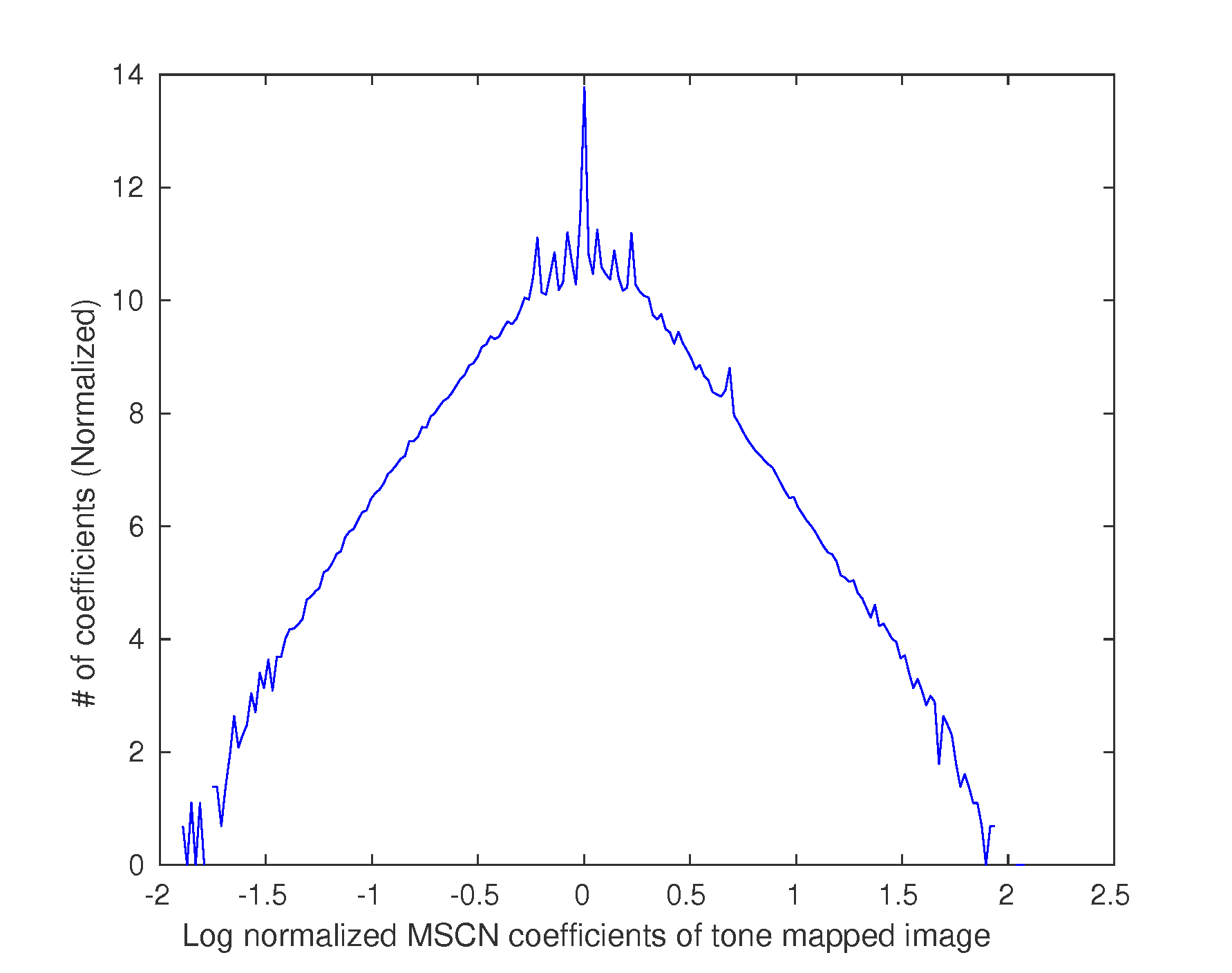}
\caption{Log normalized  MSCN coefficients distribution of the image shown 
in Fig.~\ref{fig:test_dist_tone}.}
\label{fig:hist_mscn_tone}
\end{figure}

\begin{figure}[!htbp]
\centering
\includegraphics[width=5.5cm,height=3.5cm]{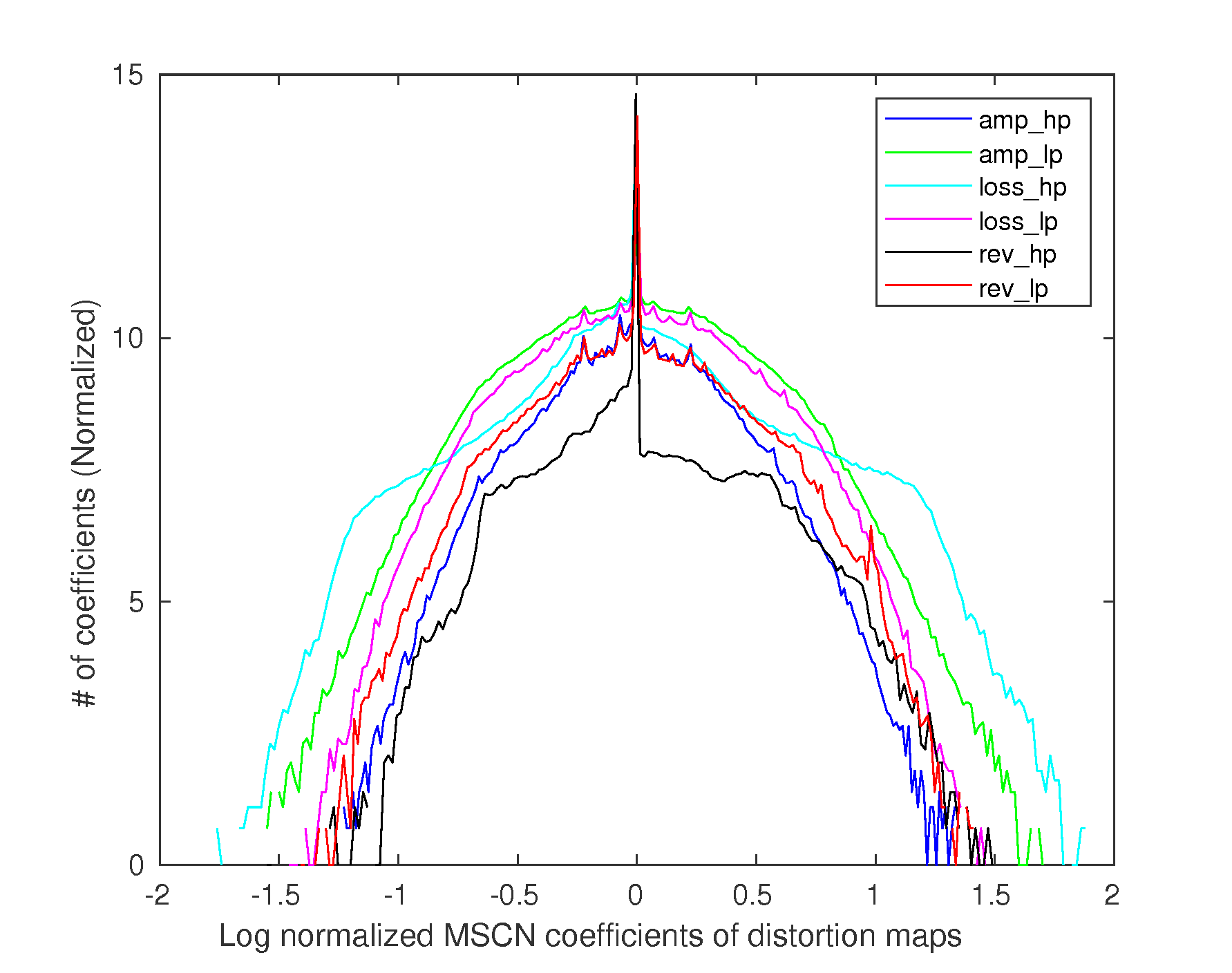}
\caption{Log normalized MSCN coefficients distribution of distortion maps of the image shown 
in Fig.~\ref{fig:test_dist_tone}.}
\label{fig:hist_mscn_dist}
\end{figure}

In Stage-I, we predicted quality aware maps to quantify the amount of distortion present at each pixel. 
To estimate the overall quality score of a tone mapped image, we require quality discerning features. We observed that the distortion maps obtained from the previous stage can be used to quantify the quality of a tone mapped image. We applied a Mean Subtraction and Contrast Normalization (MSCN) \cite{mittal2013making} transform to the tone mapped image and its corresponding distortion maps.
We observed the histograms of the resulting MSCN coefficients as shown in Fig.~\ref{fig:hist_mscn_tone} and Fig.~\ref{fig:hist_mscn_dist}. These histograms are uni-modal and are modeled using an Asymmetric Generalized Gaussian Distribution (AGGD). An AGGD with zero mean is given by:
\begin{equation}
f (x;\gamma,\beta_l,\beta_r) = 
\begin{cases}
\frac{\gamma}{ (\beta_l+\beta_r)\Gamma (\frac{1}{\gamma})} \exp \left (-\left ( \frac{-x}{\beta_l} \right)^             \gamma \right) ; \forall x < 0 \\
\frac{\gamma}{ (\beta_l+\beta_r)\Gamma (\frac{1}{\gamma})} \exp \left (-\left ( \frac{x}{\beta_r} \right)^             \gamma \right) ; \forall x \geq 0    
\end{cases}
\end{equation}
The parameters of an AGGD are estimated using the moment estimation method~\cite{lasmar2009multiscale}.
Here $\gamma > 0 $ is the shape parameter and $\beta_{l} > 0$, $\beta_{r} > 0 $ are left-scale and right-scale parameters respectively. $\Gamma (\cdot)$ is defined as,
\begin{equation}
\Gamma (a) = \int _ 0 ^ \infty t^{a-1} e^{-t} dt ; a>0
\end{equation}     

These AGGD model parameters serve as features to our regression model. We appended other statistical quantities (viz., the mean and standard deviation of the tone mapped image and its distortion maps) to the model parameters to compose the final feature vector used for supervised learning. We use support vector regression (SVR) to map these feature vectors to the quality score (mean opinion score (MOS)). 
We trained the SVR using 80\% of the data chosen randomly from our dataset. We then tested the model with the remaining 20\% of the data to measure the performance of the proposed approach. We repeated the same procedure 100 times by randomly selecting 80\% data to fit the model and the remaining 20\% data to test the model. The average of all these trials is chosen to be the final performance measures. 
\section{Results and Discussion}
\label{sec:results}
The performance of the proposed approach is evaluated and compared using the following statistical measures: Pearson linear correlation coefficient (PLCC), Spearman rank order correlation coefficient (SROCC), and root mean squared error (RMSE). The performance of the first stage was measured by comparing the predicted distortion maps with the ground truth distortion maps generated with the full reference algorithm. We can infer from Fig.~\ref{fig:compare_dist_maps} that the ground truth distortion maps and the predicted distortion maps are visually similar.

Tables~\ref{tab:compare_with_others_espl} and~\ref{tab:compare_with_others_wang} show the performance of the proposed approach on two popular HDR IQA datasets, ESPL-LIVE \cite{kundu2017large} and Yaganeh \etal~\cite{yeganeh2013objective}. Since ESPL-LIVE contains the only tone mapped images and corresponding quality (MOS) scores,  we compare our method only with existing no-reference algorithms. As we focused only on the contrast distortion we only used the tone mapped images from the ESPL-LIVE dataset \cite{kundu2017large} for performance comparison. The other dataset \cite{yeganeh2013objective} contains original HDR images, their tone mapped images, and the corresponding quality (MOS) scores. Therefore, it allows us to evaluate full reference algorithms and include them in the comparison as well. By comparing the performance of various no-reference quality assessment techniques with the proposed metric, we observe that the proposed metric shows consistent and competitive performance. We would also like to highlight that the proposed NRIQA  algorithm not only delivers competitive performance but also helps localize distortions using the distortion maps.



\begin{table}[h]
\centering
\caption{Performance comparison of various IQA algorithms on the ESPL-LIVE dataset~\cite{kundu2017large}.}
\begin{tabular}{|l|l|c|c|}
\hline
{\bf Metric}   & {\bf PLCC} & {\bf SROCC} & {\bf RMSE} \\ \hline
BIQI~\cite{moorthy2010two}       &   0.313  &   0.333    &   9.427   \\ \hline
BRISQUE~\cite{mittal2012no}       &   0.370   &    0.340   &   9.535   \\ \hline
BLIINDS-II~\cite{saad2012blind}       &   0.442   &   0.412    &  9.330    \\ \hline
DIIVINE~\cite{moorthy2011blind}        &   0.530   &    0.523   &   8.805   \\ \hline
DESIQUE~\cite{zhang2013no}        &   0.553   &   0.542    &   8.577   \\ \hline
HIGRADE-1~\cite{HIGRADE}      &   0.764   &   0.728   &   6.711   \\ \hline
HIGRADE-2~\cite{HIGRADE}       &   \textbf{0.794}   &    \textbf{0.760}   &  \textbf{6.643}    \\\hline
\textbf{RcNet (ours)}   &     0.707   &  0.667   &  7.479\\ 
\hline
\end{tabular}
\label{tab:compare_with_others_espl}
\end{table}

\begin{table}[h]
\centering
\caption{Performance comparison of various IQA algorithms on the Yeganeh dataset~\cite{yeganeh2013objective}.}
\begin{tabular}{|l|l|c|c|}
\hline
{\bf Metric}   & {\bf PLCC} & {\bf SROCC} & {\bf RMSE} \\ \hline
NIQE~\cite{mittal2013making}      &   0.368  &    0.259   &    1.788  \\ \hline
BIQI~\cite{moorthy2010two}       &   0.351   &   0.370    &   1.800  \\ \hline
ILNIQE~\cite{zhang2015feature}      &   0.326   &    0.314   &    1.818  \\ \hline
HIGRADE-1~\cite{HIGRADE}      &   0.423   &   0.270    &   1.742   \\ \hline
HIGRADE-2~\cite{HIGRADE}       &   0.563   &    0.440   &  1.589    \\ \hline
BRISQUE~\cite{mittal2012no}       &   0.569   &   0.476    &   1.581  \\ \hline
BLIINDS-II~\cite{saad2012blind}       &   0.544   &   0.508    &  1.614    \\ \hline
TMQI~\cite{yeganeh2013objective}     &  0.770   &    0.739   &    1.225  \\ \hline
\textbf{RcNet (ours)}   &     \textbf{0.847}   &  \textbf{0.768}   &  \textbf{1.029} \\ \hline
\end{tabular}
\label{tab:compare_with_others_wang}
\end{table}


\section{Conclusion}

The proposed NRIQA metric for tone mapped images has two major outcomes. The first is to localize distortions in a tone mapped image and the second is to predict its perceptual quality score in a reference free setting. Different types of distortions are visualized by what are called distortion maps. These distortion maps are learned using CNNs. We have demonstrated that these distortion maps contain quality information that can be used as features to predict the overall quality score of a tone mapped image.
We have showed that the proposed NRIQA algorithm performs competitively on two tone mapped IQA datasets.  This work can be improved at every stage by selecting and freezing robust hyper-parameters and with an efficient network architecture. Further, it is possible to add more distortion discerning features extracted from distortion maps, and select regression algorithms which are better suited to work with these features for the quality assessment task. 

\bibliographystyle{IEEEbib.bst}
\bibliography{Main}

\end{document}